\documentclass[aps,amssymb,amsmath,prl,reprint,noshowpacs]{revtex4-1}
\usepackage{times}

\usepackage{graphicx}
\usepackage[usenames]{color}

\begin{document}
	
\title{Optical potential mapping with a levitated nanoparticle at sub-wavelength distances from a membrane}

\author{Rozenn Diehl, Erik Hebestreit, Ren\'e Reimann, Martin Frimmer, Felix Tebbenjohanns and Lukas Novotny}
%\thanks{}
\affiliation{ETH Z\"urich, Photonics Laboratory, 8093 Z\"urich, Switzerland}

% To be edited by editor
% \dates{Compiled \today}

% To be edited by editor
% \doi{\url{http://dx.doi.org/10.1364/optica.XX.XXXXXX}}

\begin{abstract}
	The controllable positioning of a vacuum-levitated object near a material surface is of importance for studying short-range forces, such as Casimir forces, interfacial friction forces, or gravity in yet unexplored parameter regimes. Here we optically levitate a nanoparticle in a laser beam strongly focused on a dielectric membrane. By investigating the motion of the trapped particle in vacuum, we map the position-dependent optical potential of the particle. We interferometrically measure the distance between the particle and the surface and demonstrate stable trapping in sub-wavelength proximity of the dielectric surface. Our work is important for the development of on-chip levitated optomechanics and for measuring short-range forces at sub-wavelength distances.
\end{abstract}

\maketitle

\section{Introduction}

Levitated optomechanical systems are prime candidates for future optomechanical sensing technologies operating at the limits set by our current understanding of physics~\cite{YIN2013,Romero-Isart2010}. The position of an optically levitated nanoparticle can be measured with a precision ultimately governed by the Heisenberg uncertainty relation of quantum mechanics~\cite{Aspelmeyer2014}. This measurement precision has been harnessed to control the levitated particle's center-of-mass dynamics and feedback-cool its motion towards the quantum ground state~\cite{Li2011,Gieseler2012,Kiesel2013,Millen2015,Jain2016}. Furthermore, this precision is at the heart of optically levitated sensing systems for static~\cite{Hebestreit2017}, resonant~\cite{Ranjit2016,Hempston2017}, and inertial forces~\cite{Armano2016,Monteiro2017}.
Currently, levitated optomechanics is making the critical transition from a laboratory system under investigation from a fundamental perspective towards a technology for applied research and engineering, which requires on-chip integration of optically levitated nanoparticles~\cite{Alda2016,Schein2015,Schmidt2007}. An essential step towards exploiting the synergies of levitated optomechanics and integrated optics is to controllably position an optically levitated nano-object in sub-wavelength proximity of a dielectric surface. This has recently been demonstrated for a single atom in the vicinity of photonic crystals~\cite{Tiecke2014}.
Reaching beyond technological applications, due to its outstanding force sensitivity, an optically levitated nanoparticle at a nanometric distance from an interface can provide the sensitivity to investigate short-range forces~\cite{Geraci2010,Arvanitaki2013,Schein2015,Winstone2017} and mesoscopic thermodynamics~\cite{Tschikin2012,Gieseler2014}.
Despite the tantalizing opportunities for applications and fundamental research, optical levitation of a nanoparticle at a sub-wavelength distance from an interface has remained elusive to date.

In this paper, we map the optical potential generated by a laser beam strongly focused onto a dielectric membrane. To this end, we measure the center-of-mass motion of a dielectric nanoparticle optically trapped in vacuum in the focused field. As we approach the dielectric membrane to the laser focus, the oscillation frequency of the particle serves as a measure for the varying trap stiffness resulting from interference of the trapping light with its reflection from the membrane. We simultaneously determine the distance between the particle and the membrane using an interferometric technique. Our approach allows us to controllably reduce this distance to below half of the wavelength of the trapping laser.

\section{Method}
%\label{sec:examples}

\begin{figure}[h]
	\includegraphics[width=\linewidth]{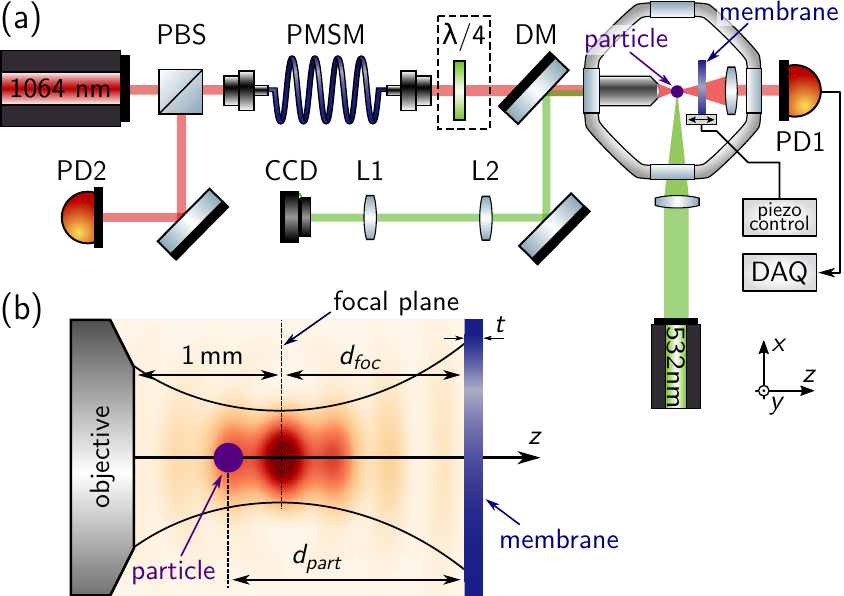}
	\caption{(a)~Experimental setup. A silica nanoparticle is trapped inside a vacuum chamber in the focus of a laser beam (1064~nm) which is spatially filtered by a polarization maintaining single-mode fiber (PMSM). The forward-scattered light from the nanoparticle is detected by a balanced photodetector (PD1). A SiN membrane (thickness 500~nm) is placed between the particle and the collection lens on a piezoelectric stage. To calibrate the focus-to-surface distance $d_\text{foc}$, we introduce a quarter waveplate ($\lambda/4$) such that the light backreflected from the surface is deflected by the polarizing beam splitter (PBS), and can be detected on a photodetector (PD2). To determine the particle-to-surface distance $d_\text{part}$, a linearly polarized green laser (532~nm, 100~mW) is weakly focused on the nanoparticle and the back focal plane of the objective is imaged onto a camera (CCD) using lenses L1 and L2 and the dichroic mirror (DM). (b)~Illustration of relevant geometrical parameters (not to scale). The distance $d_\text{foc}$ between the focal plane of the objective and the front surface of the SiN membrane (nominal thickness $t=500~$nm) can be adjusted by moving the membrane with the piezostage. The optical potential (illustrated by the red intensity distribution) results from focused laser beam interfering with its own reflection at the membrane and determines the distance $d_\text{part}$ between the particle and the membrane surface.}
	\label{setup}
\end{figure}

Our experimental setup is shown in Fig.~\ref{setup}(a). A linearly polarized laser beam (wavelength 1064~nm, focal power 100~mW) is focused by an objective (100$\times$, 0.9NA) to trap a single silica nanoparticle of $136\,\text{nm}$ diameter. A collection lens collimates the light scattered by the particle in the forward direction. We use the balanced homodyne detection scheme (PD1) detailed in Ref.~\cite{Gieseler2012} to measure the particle position along all three axes ($z$ denotes the optical axis and $x$ the polarization direction). Between particle and collection lens, we place a suspended silicon nitride (SiN) membrane (thickness $t=500~$nm) transverse to the optical axis. The membrane is mounted on a piezoelectric stage which allows us to position the membrane with nm precision relative to the focal plane of the objective.  Prior to any measurement, we discharge the particle to avoid undesired electrostatic interactions with the membrane~\cite{Brownnutt2015,Frimmer2017}.

The relevant geometrical parameters of our experiment are illustrated in Fig.~\ref{setup}(b). The distance between the objective and the focal plane is fixed to 1~mm by the working distance of our objective. By moving the membrane along the optical axis, we can adjust the distance $d_\text{foc}$ between the membrane's front surface and the focal plane. Importantly, back reflections from the membrane modify the optical potential, such that the trapped particle does not necessarily reside in the focal plane of the objective. We denote the distance between the particle and the membrane as $d_\text{part}$.

To calibrate the distance $d_\text{foc}$ between the membrane surface and the objective's focal point, we introduce a quarter-waveplate ($\lambda/4$) in the beam path before the objective. Without a particle in the trap, we scan the membrane position along the optical axis through the focal region while detecting the backreflected intensity on a photodiode (PD2). A polarization maintaining single-mode fiber (PMSM) acts as a spatial filter, such that the signal on PD2 peaks when the membrane surface coincides with the focal plane and the focus of the objective is exactly imaged onto the fiber.

\section{Results and discussion}
\subsection{Measurement of trapping potential}
\begin{figure}[h]
	\includegraphics{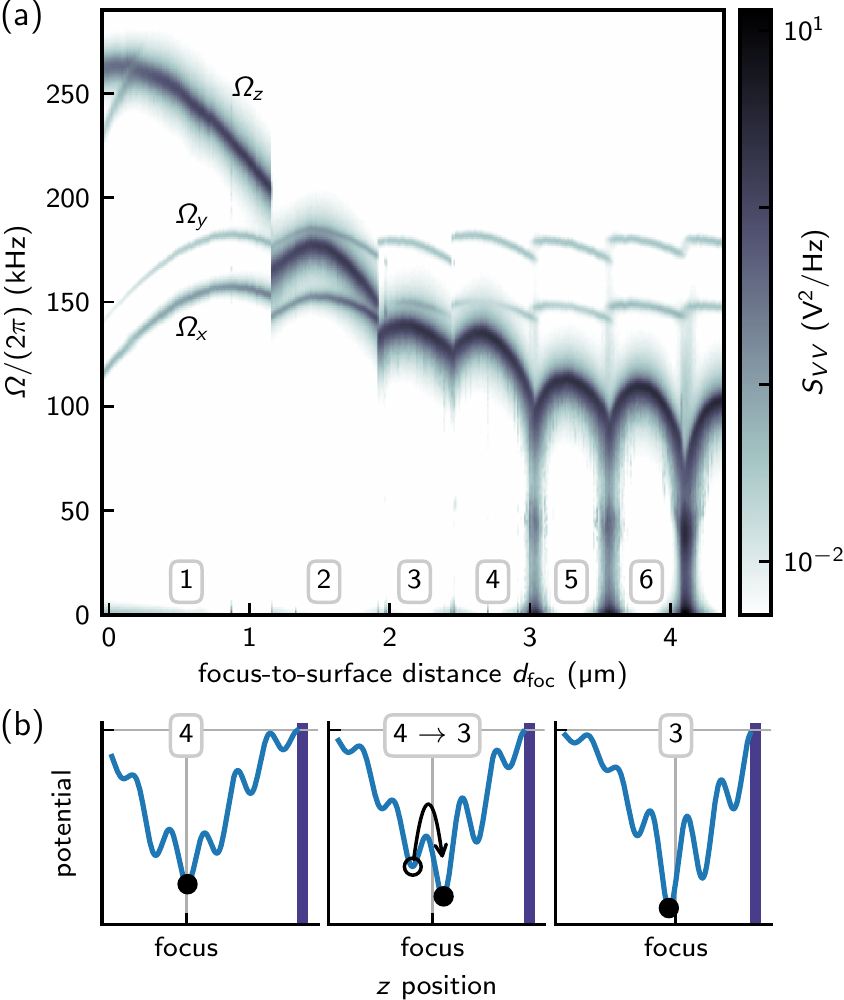}
	\caption{(a) Measured power spectral density of the center-of-mass motion of the nanoparticle, for varying focus-to-surface distances $d_\text{foc}$. The strong feature in each spectrum is the eigenmode  of the particle's center-of-mass motion along the $z$ axis with frequency $\Omega_z$. The two weaker features at $\Omega_x$ and $\Omega_y$ correspond to oscillations along the transverse directions (parallel to the surface), at around 150~kHz and 175~kHz, respectively.
		(b) Sketches of the optical potential along the optical axis illustrating the particle's transition from potential well~4 to well~3. The wells are generated by interference of the incoming trapping laser with its reflection at the membrane. We number the wells starting with the one closest to the interface. Left: The particle is in well~4. Middle: After approaching the surface, the potential barrier is small enough for the particle to transition from well~4 to well~3. Right: The particle is now in potential well~3.
		\label{travelling}}
\end{figure}

In order to investigate the optical trapping potential in close proximity of the SiN membrane, we measure the particle dynamics as a function of the distance between the membrane and the focal plane.
To this end, at a pressure of 1.5~mbar, we move the membrane towards the objective at a speed of 20~nm/s. During this approach, we record 600~ms-long time traces of the particle's motion on detector PD1 with a sampling rate of 667~kHz. From each time trace, we calculate the power spectral density $S_{VV}(\Omega)$~\cite{Clerk2010} and display the result in Fig.~\ref{travelling}(a) as a false-color plot with $d_\text{foc}$ on the horizontal, and frequency $\Omega$ on the vertical axis.
We observe salient dark regions in Fig.~\ref{travelling}(a), corresponding to a strong peak in the power spectrum, which undergoes periodic frequency shifts between 50 and 150~kHz as the focus-to-surface distance $d_\text{foc}$ is reduced down to about $2~\mu$m. We attribute this strong peak to the particle's oscillation mode along the optical axis $z$. For even smaller values of $d_\text{foc}$, this strong peak shows an overall shift towards higher frequencies, exceeding 250~kHz as $d_\text{foc}$ goes towards zero. Notably, at certain values of $d_\text{foc}$, the peak position makes a discrete jump (\emph{e.g.}, at $d_\text{foc}=1.1~\mu$m). Furthermore, we observe two less pronounced features [around $\Omega_x/(2\pi)=150~$kHz and $\Omega_y/(2\pi)=170~$kHz for $d_\text{foc}=4~\mu$m], which correspond to the motion of the $x$ and $y$ modes, showing up on the $z$ detector due to inevitably imperfect alignment. The oscillation frequency of the $y$ mode exceeds that of the $x$ mode due to the weaker confinement of the optical focus along the polarization direction of the field. All modes are thermally populated by collisions of the particle with gas molecules in the vacuum chamber. Importantly, the mode frequencies $\Omega_x,\Omega_y,\Omega_z$ are directly proportional to the curvature of the trapping potential confining the particle.

We can intuitively understand the experimentally observed variation of the trap frequency with the approaching membrane by considering the intensity distribution of an optical beam strongly focused onto a dielectric membrane~\cite{Novotny2006}. The interaction of the trapped particle with the optical field is dominated by the gradient force. The interference of the incoming trapping beam with its back reflection from the membrane generates a standing-wave in front of the membrane. Field intensity $I$ and optical potential are linked by the equation $U=-\alpha^\prime I /(2c\varepsilon_0)$~\cite{Gieseler2012}, where $\alpha^\prime = 9.39\times 10^{-33}\,\mathrm{F\cdot m^{2}}$ is the real part of the particle's polarizability, $c$ is the speed of light and $\varepsilon_0$ is the vacuum permittivity. These back reflections therefore modulate the trapping potential. As a result, due to the presence of the membrane, several local optical potential minima form, which we number starting with the first intensity maximum closest to the membrane. In Fig.~\ref{travelling}(b), we illustrate the typical behavior of both the optical potential and the particle during the approach of the membrane. The particle resides in one potential minimum whose curvature depends on the exact membrane position and determines the oscillation frequency of the particle. In our example, the particle is in well~4 (left). As the membrane is moved closer to the focal plane, the standing wave pattern is shifted through the region of strongest lateral confinement of the beam and the particle is moved along with the standing wave pattern towards the objective. With the membrane moving closer, the barrier between the adjacent well becomes small enough compared to the thermal energy for the particle to transition into the next well towards the membrane. This transition is depicted in Fig.~\ref{travelling}(b, middle) where the particle transitions from well~4 to well~3, which becomes the global minimum as the membrane approaches even further [Fig.~\ref{travelling}(b, right)].
Accordingly, as we approach the membrane further, the particle is ``handed'' from well to well, leading to the periodic modulation of its oscillation frequencies observed in Fig.~\ref{travelling}(a). The boxed numbers in Fig.~\ref{travelling} indicate the number of the well the particle resides in as the membrane approaches. 
The strong increase of the oscillation frequency $\Omega_z$ for small distances $d_\text{foc}$ is explained by the stronger localization of the reflected field as the membrane enters the focal region of the field. This explanation is supported by the fact that the length scale on which the frequency $\Omega_z$ is strongly increased corresponds to the Rayleigh length of the laser focus, which is around $1~\mu$m. 
In addition, we note that $\Omega_x$ and $\Omega_y$ decrease significantly for $d_\text{foc}$ smaller than $0.8~\mu$m. We explain this behavior by the widening of the potential along the transverse directions as well~1 is pushed away from the geometrical focus.
Finally, we note the spectral feature at 250~kHz crossing the $z$ mode around $d_\text{foc}\approx0.2~\mu$m, which is the second harmonic of $\Omega_x$ and weakly couples to the $z$ motion, which can be explained by non-linearities of the optical potential sampled by the particle at room temperature~\cite{Gieseler2013}.

\begin{figure*}%[t!]
	\includegraphics[width=2\columnwidth]{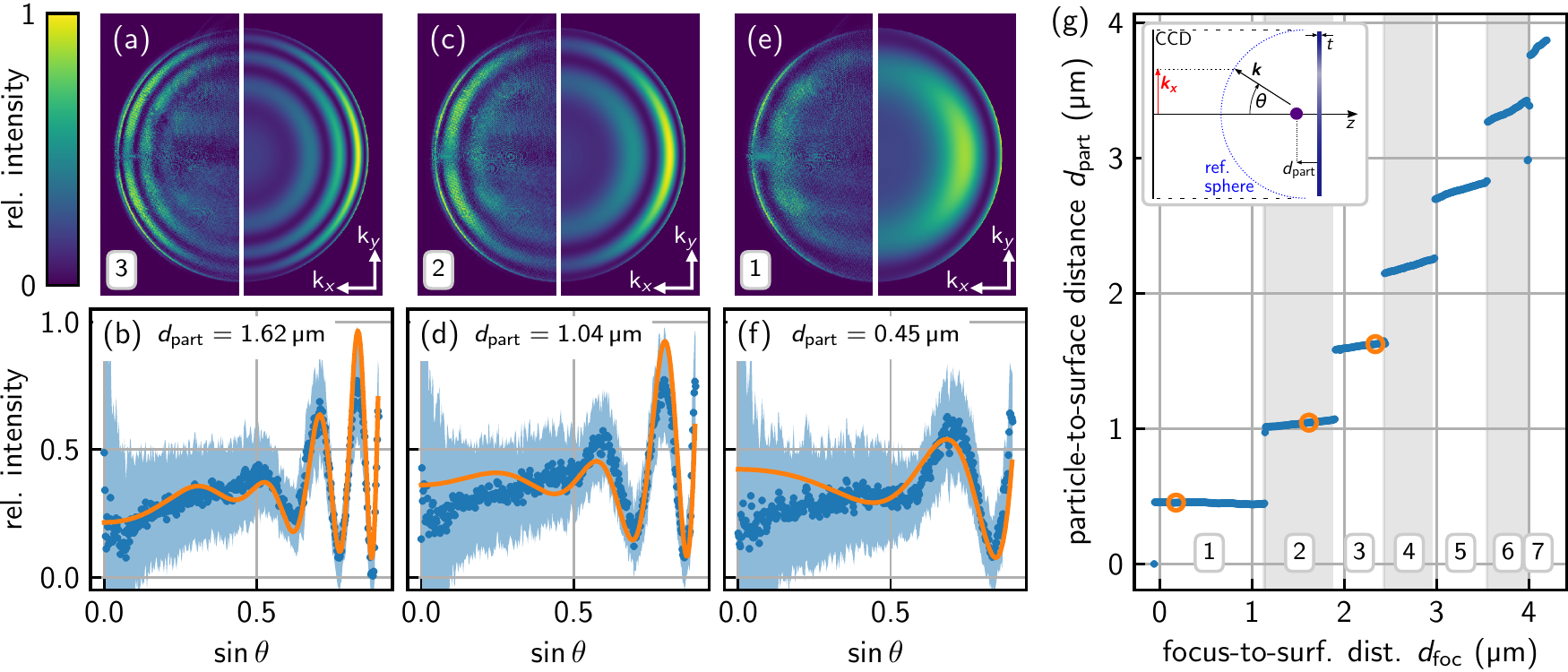}
	\caption{Back-focal-plane imaging of a nanoparticle levitated in front of a SiN membrane ($k_x$ and $k_y$ denote the in-plane components of the wavevector). The scattering angle $\theta$ relative to the optical axis is given by $k\text{sin}(\theta) = \sqrt{k_x^2 + k_y^2}$ [see inset in (g)]. (a, left-half)~Image of the back focal plane measured at $d_\text{foc}=2.34~\mu\text{m}$. (right-half) Calculated back focal plane intensity distribution for a scatterer levitated at a distance $d_\text{part}=1.62~\mu\text{m}$ [extracted from the fit in (b)]. (b)~Radially averaged intensity profile (blue points) from the measured image in (a). $\text{sin}(\theta)$ is defined such that $\text{sin}(\theta_\text{max}) = \text{NA} = 0.9$. The blue shaded area corresponds to the experimental standard deviation. We fit (orange curve) the intensity profile according to our model, from which we extract the distance between the scatterer and the reflecting surface $d_\text{part}=1.62~\mu\text{m}$. (c), (d) and (e), (f) analog to (a), (b) but for images recorded in areas 2 and 1, respectively [see Fig.~\ref{travelling}(a)]. (g)~Plot of the focus-to-surface distance $d_\text{foc}$ versus the particle-to-surface distance $d_\text{part}$. The values for $d_\text{part}$ are obtained by fitting the back focal plane images recorded during the entire approach of the membrane. The orange circles indicate the three examples detailed in Figs.~\ref{rings}(a,c,e).
		\label{rings}}
\end{figure*}

\subsection{Interferometric measurement of particle-to-surface distance}

Thus far, we have characterized the potential landscape in front of the dielectric membrane as a function of membrane position, measured relative to the focal plane of the objective. We now turn to a measurement of the distance $d_\text{part}$ between the levitated nanoparticle and the membrane surface. For this measurement, we illuminate the particle from the side with a linearly polarized green laser field at 532~nm [see Fig.~\ref{setup}(a)]. The field collected by the objective corresponds to the superposition of the particle's scattered field and the particle's scattered field after it has been reflected by the membrane, similar to spectral self-interference microscopy~\cite{Davis2007} The phase difference between these two fields depends on the radiation angle and leads to a characteristic interference pattern in the back focal plane of the objective. This interference pattern provides a means to gauge the distance between the particle and the interface. With a dichroic mirror [DM in Fig.~\ref{setup}(a)], we separate the green light from the trapping light and image the back focal plane of the objective onto a camera. In Fig.~\ref{rings}(a, left), we show a back focal plane image recorded for $d_\text{foc}=2.34~\mu$m, where the particle is in well 3. We observe the dipolar radiation pattern of the point scatterer, modulated with a characteristic ring pattern, which encodes the distance between the particle and the surface $d_\text{part}$. The distance $d_\text{part}$ corresponds roughly to the number of fringes (interference undulations) multiplied by $\lambda/2$. To extract $d_\text{part}$ more accurately, we radially average the measured intensity distribution in the back focal plane [blue data points in Fig.~\ref{rings}(b)] and fit the resulting intensity profile with the theoretical model described in Ref.~\cite{Novotny2006b}. For our calculation, we take into account the thermal motion of the particle in the optical potential by performing a correspondingly weighted average of the back focal plane images for the values of $d_\text{part}$ sampled by the particle. We obtain the best fit to our measurement for $d_\text{part} = (1.62 \pm 0.05)~\mu\text{m}$, shown as the orange line in Fig.~\ref{rings}(b). We estimate the error on $d_\text{part}$ based on the residuals from the fit. For visual comparison, we show the corresponding theoretical back focal plane image in the right half of Fig.~\ref{rings}(a). 

We show two more back focal plane images for $d_\text{foc}=1.62~\mu$m [Fig.~\ref{rings}(c), particle in well~2] and $d_\text{foc}=0.18~\mu$m [Fig.~\ref{rings}(d), particle in well~1]. Our analysis of the averaged intensity profiles shown in Fig.~\ref{rings}(d) and (f) yields the values $d_\text{part} = (1.04 \pm 0.05)~\mu\text{m}$ (well~2) and $(450 \pm 50)~\text{nm}$ (well~1), respectively.

We note that besides the overall power, and the particle-to-surface distance $d_\text{part}$, the only free parameter of the fit is the membrane thickness $t$. We consistently extract a value $t = (516~\pm 1)~\text{nm}$, which lies within the tolerance range provided by the manufacturer ($(500~\pm 25)~\text{nm}$).

With our interferometric measurement technique, by collecting a back focal plane image at every position $d_\text{foc}$ of the approaching membrane, we can continuously monitor the particle-to-surface distance $d_\text{part}$, as shown in Fig.~\ref{rings}(g). The orange circles indicate the positions at which the 3 images displayed in (a), (c), and (e) were recorded. We note that the discrete steps in $d_\text{part}$ occur together with the discontinuities of the oscillation frequencies observed in Fig.~\ref{travelling}(a), since both effects arise from the fact that the particle transitions into a neighboring well. This explanation is further supported by the observation that the step size in $d_\text{part}$ observed in Fig.~\ref{rings}(g) is approximately $\lambda/2$, which corresponds to the spacing of the intensity maxima in a standing wave. We furthermore note that this step size increases for small $d_\text{foc}$, a feature that we attribute to the additional Gouy phase that is acquired near the focus.
The separation of well~1 from the membrane surface is 450~nm. This value corresponds to less than half of the trapping laser wavelength and is set by the complex reflection coefficient of the dielectric membrane and depends on its thickness and refractive index.

\section{Conclusion}

We have optically levitated a silica nanoparticle at a sub-wavelength distance of 450~nm from a SiN membrane. By analyzing the particle's motion, we have characterized the trapping potential in close proximity to the dielectric interface. With an independent interferometric technique, we have measured the exact distance between the particle and the surface of the membrane, providing the precise location of our probe during the mapping of the potential. Our experiments mark an important step towards integrated devices based on the combination of levitated optomechanics with optical waveguiding structures~\cite{Schein2015,OShea2013,Daly2016,Schmidt2007}.
Furthermore, the controlled optical levitation of a nanoparticle in close proximity of an interface is an integral requirement to harness the outstanding force sensitivity of levitated optomechanics for the characterization of weak short-range interactions, such as Casimir forces, non-contact friction, and ultimately non-Newtonian gravity in yet unexplored parameter regimes~\cite{Geraci2010,Almasi2015,Garrett2018}. Operating with uncharged particles, our system is insensitive to surrounding electrostatic fields, which are usually orders of magnitude stronger than surface forces. Electrostatic forces would limit our sensitivity and therefore hinder the measurement of surface forces. Finally, our technique brings measurements of thermal transfer dynamics at the nanoscale within reach~\cite{Tschikin2012}.

\section*{Funding Information}
This research was supported by ERCQMES (Grant No. 338763) and the NCCR-QSIT program (Grant No. 51NF40-160591).

\section*{Acknowledgments}

The authors thank D. Windey and J. Gieseler for fruitful scientific discussions.

%\section*{Supplemental Documents}

%\bigskip \noindent See \href{link}{Supplement 1} for supporting content.

%Note that letter submissions to \emph{Optica} use an abbreviated reference style. Citations to journal articles should omit the article title and final page number; this abbreviated reference style is produced automatically when the \texttt{$\setminus$setboolean\{shortarticle\}\{true\}} option is selected in the template, if you are using a .bib file for your references. 
%
%However, full references (to aid the editor and reviewers) must be included as well on an informational page that will not count against page length; again this will be produced automatically if you are using a .bib file and have the \texttt{$\setminus$setboolean\{shortarticle\}\{true\}} option selected.

% Bibliography
%\bibliography{biblio_paper_c}

% Full bibliography will be added automatically on a new page for Optics Letters submissions. This command is ignored for journal article submissions.
% Note that this extra page will not count against page length.
%\bibliographyfullrefs{sample}

\end{document}